# Sequential sample size calculations and learning curves safeguard the robust development of a clinical prediction model for individuals

Amardeep Legha[1,2]*, Joie Ensor[1,2], Rebecca Whittle[1,2], Lucinda Archer[1,2,3], Ben Van Calster[4,5], Evangelia Christodoulou[6], Kym I.E. Snell[1,2], Mohsen Sadatsafavi[7], Gary S. Collins[8], Richard D. Riley[1,2]

**Author details:**

* Corresponding author: Amardeep Legha; a.legha@bham.ac.uk

[1] Department of Applied Health Sciences, School of Health Sciences, College of Medicine and Health, University of Birmingham, Birmingham, United Kingdom.
[2] National Institute for Health and Care Research (NIHR) Birmingham Biomedical Research Centre, United Kingdom.
[3] Institute of Data and AI, University of Birmingham, United Kingdom.
[4] Leuven Unit for Health Technology Assessment Research (LUHTAR), KU Leuven, Leuven, Belgium
[5] Department of Development and Regeneration, KU Leuven, Leuven, Belgium
[6] German Cancer Research Center (DKFZ) Heidelberg, Division of Intelligent Medical Systems, Heidelberg, Germany
[7] Respiratory Evaluation Sciences Program, Faculty of Pharmaceutical Sciences, University of British Columbia, Vancouver, Canada
[8] Centre for Statistics in Medicine, Nuffield Department of Orthopaedics, Rheumatology and Musculoskeletal Sciences, University of Oxford, Oxford, United Kingdom.

**Funding:** This paper presents independent research supported by an EPSRC grant for 'Artificial intelligence innovation to accelerate health research' (number: EP/Y018516/1), the MRC Better Methods Better Research panel (grant reference: MR/Z503873/1) and by the National Institute for Health and Care Research (NIHR) Birmingham Biomedical Research Centre at the University Hospitals Birmingham NHS Foundation Trust and the University of Birmingham. GSC is supported by Cancer Research UK (programme grant: C49297/A27294). RDR and GSC are National Institute for Health and Care Research (NIHR) Senior Investigators. The views expressed are those of the author(s) and not necessarily those of the NHS, the NIHR or the Department of Health and Social Care.

**Competing Interests:** RDR receives royalties for sales of his textbooks "Prognosis Research in Healthcare" and "Individual Participant Data Meta-analysis".

**Acknowledgements:** Not applicable

**Ethics approval and consent to participate:** Not applicable

**Data Availability:** The dataset was obtained from the Medical Information Mart for Intensive Care III (MIMIC-III) trial, which contains freely available and de-identified critical care data from the Beth Israel Deaconess Medical Center in Boston, Massachusetts, between 2001 and 2012. See: Johnson AEW, Pollard TJ, Shen L, Lehman L-wH, Feng M, Ghassemi M, et al. MIMIC-III, a freely accessible critical care database. Scientific Data. 2016;3(1):160035.

# CRediT author statement:

Amardeep Legha: Methodology, Software, Formal analysis, Writing - Original Draft, Writing - Review & Editing, Visualization

Joie Ensor: Methodology, Software, Writing - Review & Editing, Supervision.

Rebecca Whittle: Methodology, Software, Writing - Review & Editing.

Lucinda Archer: Software, Writing - Review & Editing, Visualization.

Ben Van Calster: Methodology, Writing - Review & Editing.

Evangelia Christodoulou: Methodology, Writing - Review & Editing.

Kym I.E. Snell: Writing - Review & Editing.

Mohsen Sadatsafavi: Methodology, Writing - Review & Editing.

Gary S. Collins: Methodology, Writing - Review & Editing.

Richard D. Riley: Conceptualization, Methodology, Software, Resources, Writing - Original Draft, Writing - Review & Editing, Supervision, Funding acquisition.

# Abstract

**Background:**

When recruiting participants to a new study developing a clinical prediction model (CPM), sample size calculations are typically conducted before data collection based on sensible assumptions. This leads to a fixed sample size, but if the assumptions are inaccurate, the actual sample size required to develop a reliable model may be very different. To safeguard against this, adaptive sample size approaches have been proposed, based on sequential evaluation of (changes in) a model's predictive performance.

**Objectives:**

To illustrate and extend sequential sample size calculations for CPM development by (i) proposing stopping rules for prospective data collection based on minimising uncertainty (instability) and misclassification of individual-level predictions, and (ii) showcasing how it safeguards against inaccurate fixed sample size calculations.

**Methods:**

Using the sequential approach repeats the pre-defined model development strategy every time a chosen number (e.g., 100) of participants are recruited and adequately followed up. At each stage, CPM performance is evaluated using bootstrapping, leading to prediction and classification stability statistics and plots, alongside optimism-adjusted measures of calibration and discrimination. Learning curves display the trend of results against sample size and recruitment is stopped when a chosen stopping rule is met.

**Results:**

Our approach is illustrated for model development of acute kidney injury using (penalised) logistic regression CPMs. Prior to recruitment based on perceived sensible assumptions, the fixed sample size calculation suggests recruiting 342 patients to minimise overfitting; however, during data collection the sequential approach reveals that a much larger sample size of 1100 is required to minimise overfitting (targeting a bootstrap-corrected calibration slope ≥0.9). If the stopping rule criteria also target small uncertainty and misclassification probability of individual predictions, the sequential approach suggests an even larger sample size of about *n*=1800.

**Conclusions:**

For CPM development studies involving prospective data collection, a sequential sample size approach allows users to dynamically monitor individual-level prediction and classification instability. This helps determine when enough participants have been recruited and safeguards against using inaccurate assumptions in a sample size calculation prior to data recruitment. Engagement with patients and other stakeholders is crucial to identify sensible context-specific stopping rules for robust individual predictions.

**Abstract Word Count:** 348 words

**Manuscript Word Count (excluding Figures/Tables/Supplementary Material/References):** 3,409 words

**Highlights:**

- Researchers must ensure adequate sample sizes to develop clinical prediction models
- During prospective data collection, sequential sample size calculations and learning curves safeguard robust model development
- Stopping rules for new participant recruitment can be based on (changes in) individual-level prediction and classification instability
- The sequential approach may identify very different required sample sizes than a priori calculations based on inaccurate assumptions
- Engagement with key stakeholders is crucial to identify sensible stopping rules

**What is new?**

Key findings

- When sequentially determining a minimum sample size for prospectively developing a prediction model, stopping rules based on minimising uncertainty (instability) and misclassification of individual-level predictions suggest much larger sample sizes than stopping rules based only on achieving stable population-level performance alone
- Safeguarding against making inaccurate assumptions in a sample size calculation prior to data recruitment is important as the actual sample size required to develop a robust model may be very different – the sequential sample size approach proposed allows for this safeguarding of robust model development

What this adds to what is known?

- Sequential sample size calculations have been previously applied to ensure population-level stability; this new work shows how to extend to examine individual-level stability and why it safeguards against inaccurate assumptions used in sample size calculations before data collection

What is the implication/what should change now?

- An adaptive approach to sample size calculations should be considered when designing a prospective study to develop a clinical prediction model to ensure the developed model is robust
- Particular consideration should be given towards making predictions stable at the individual-level. Key stopping rule criteria and acceptable levels of individual-level stability is to be determined by engaging with key stakeholders

# 1. Introduction

Clinical prediction models (CPMs) map an individual's characteristics (predictors) to an estimated outcome value or the risk of a particular outcome occurring. CPMs are common in the medical literature. For example, Wynants et al (2020)[1] identified 381 newly developed prognostic CPMs for COVID-19 published within the first year of the pandemic. However, the authors concluded that most of these CPMs were poorly reported with methodological flaws and were at high risk of bias, which limits the suitability of such models for clinical decision making. Similar issues have been observed in other reviews of CPMs developed using machine learning approaches.[2-4]

A key criterion when reporting CPMs is to explain how the study sample size was arrived at and justify it is sufficient to answer the research question, as advocated by TRIPOD+AI (2024; Transparent Reporting of a multivariable model for Individual Prognosis Or Diagnosis + Artificial Intelligence).[5] An adequate sample size is needed when developing a CPM to reduce the risk of model overfitting, ensure stability, and achieve good calibration in the target population.[6] Comprehensive methods for calculating minimum sample size requirements have been proposed by Riley et al (2019a; 2019b; 2020),[7-9] and can be implemented using the R, Stata, or Python module *pmsampsize*.[10, 11] These methods target precise estimation of the overall event risk, minimal overfitting of predictor effects, and small optimism in the apparent model fit; thus targeting a CPM that is fit-for-purpose at least at the population-level.

However, as shown more recently by Riley and Collins (2023),[12] it is also important to ensure that a CPM has stable performance (in terms of model calibration, discrimination, and clinical utility) at the individual level. Ultimately, healthcare professionals will use CPMs to inform clinical decisions for an individual patient, not for an entire population, thus CPMs with high uncertainty at the individual-level may be inappropriate for clinical use.[13] Recent sample size guidance aims to examine individual-level stability in predictions, to guide the sample size needed in advance of data collection or to decide if an existing dataset is suitable for model development.[14, 15]

Currently, sample size calculations are typically performed prior to data collection, based on assumptions about key parameters such as the model's c-statistic and the overall risk in the population.[7-9] However, these assumptions may not hold once data are obtained and the model is actually developed, meaning the predetermined sample size may be insufficient for the intended analyses. An alternative is to use an adaptive or sequential approach to sample size calculations, where the target sample size is regularly updated as new patient data are collected, to identify when sufficient data have been collected, based on pre-specified stopping rules. This may lead to more reliable CPMs being developed than when using a fixed sample size approach determined a priori.

To address this, Christodoulou et al (2021)[16] proposed an adaptive sample size approach for developing a CPM during prospective data collection. The approach sequentially checks and plots (via learning curves) the robustness of a model's predictions as new data are obtained, for example in terms of (changes in) overfitting, optimism and performance. This is then used to inform stopping rules for when to stop new data collection. The authors demonstrate through case studies that potentially larger sample sizes were needed compared to the fixed sample size method of Riley et al (2019)[8] to achieve stability in the learning curve for population-level measures such as the (optimism in the) calibration slope and c-statistic. Nevertheless, individual-level stability was not examined and so a concern is that, given the findings of by Riley and Collins (2023),[12] the sequential sample size approach may identify that even larger sample sizes are needed to ensure stable individual-level predictions and classifications.

To address this, in this paper we illustrate and extend sequential sample size calculations for CPM development by (i) proposing stopping rules for prospective data collection based on minimising uncertainty (instability) and misclassification of individual-level predictions, and (ii) showcasing how it safeguards against using inaccurate sample sizes determined a priori (i.e., before data collection) based on wrong assumptions.[17] We illustrate the approach in examples using various statistical development approaches and compare the resulting sample size requirements to those based on existing criteria.[8] We then provide discussion and make recommendations for the field.

## 2. Methods

In this Section, we introduce a motivating example and outline the proposed sequential sample size calculation approach, with extension to examine individual-level prediction stability.

## 2.1 Motivating Example

We use data from the Medical Information Mart for Intensive Care III (MIMIC-III) database[18] to illustrate our proposed approach. MIMIC-III is a freely available database, containing over 40,000 patients admitted to the intensive care unit (ICU) of the Beth Israel Deaconess Medical Center in Boston, Massachusetts, between 2001 and 2012. From these, we selected a prospective cohort of 20,413 adult patients (aged ≥18 years) who were admitted to the ICU for any reason for at least 24 hours. The outcome of interest was the development of acute kidney injury (AKI), defined as present when the maximum creatinine level within 48 hours of the end of the first day on ICU was greater than the day one minimum creatinine level by either: (i) > 1.5 times, or (ii) > 0.3 mg/dL.[19] The prevalence of AKI was 17.3% in this dataset. To mirror prospective data collection to a new cohort study, we randomly assigned a recruitment order to participants, from 1 (first) to 20,413 (last).

Six core predictor parameters were used to illustrate our method, taken from a previous clinically-related study by Zimmerman et al (2019):[20] bicarbonate (mg/dL), creatinine (mg/dL), haemoglobin (g/dL), blood urea nitrogen (mg/dL), potassium (mg/dL), and systolic blood pressure (mmHg). These six predictors were all continuous and linear trends were assumed between the predictors and risk of AKI for simplicity, though in practice non-linear trends could also be investigated.

## 2.2 Sequential Process to Examine Sample Size and Generate a Learning Curve

When undertaking prospective data collection (i.e., recruitment and follow-up of individuals) for a study developing a new CPM, the following process can be used to sequentially examine sample size requirements and to generate a learning curve for the model performance. The learning curve shows how estimates of a chosen performance statistic change as the sample sizes increases incrementally and can be used to help decide when to stop new participant recruitment. The process (Figure 1) adapts the approach of Christodoulou et al (2021)[16] by extending to consider (via Harrell's bootstrap[21]) individual-level stability of predictions (see step (v)).

**Figure 1.** Sequential Process to Generate a Learning Curve for Examining the Impact of Sample Size on Prediction Model Performance and Stability of Predictions

i) Begin by recruiting an initial sample of patients $N_{initial}$. This forms the initial model development dataset.

ii) Apply a pre-specified model building strategy to the initial model development dataset (see Section 2.4 for examples of different possible modelling strategies). Estimate and store the model's predictions $\hat{p}_{initial\_i}$ for each individual ($i = 1$ to $N_{initial}$), as well as the value of the performance statistic of interest, $\Theta_{initial}$ (e.g., measures of calibration, discrimination or clinical utility; see Section 2.3 for details of these measures).

iii) Apply Harrell's bootstrap method[21] using $B$ bootstrap samples, as follows:
   a) Draw a bootstrap sample of size $N_{initial}$ (with replacement) from the initial model development dataset.
   b) Apply the modelling strategy used in Step (ii) (including the approach for any parameter tuning or data splitting) to the bootstrap sample, to obtain a bootstrap model $M_b$, where $b$ represents the bootstrap sample used to generate the model ($b = 1$ to $B$). Calculate and store the estimated performance measures of the bootstrap model, $M_b$, when applied in bootstrap sample $b$ ($\Theta_b$).
   c) Apply the bootstrap model $M_b$ to estimate and store the risk ($\hat{p}_{initial\_bi}$) for each individual $i$ in the original development dataset ($N_{initial}$) from Step (i), and estimate and store the model performance measures $\Theta_{initial\_b}$ in that data.
   d) Calculate the optimism in model performance as the difference in the bootstrap model's apparent (in bootstrap sample) and test (in original data) performance, i.e., $\Theta_b - \Theta_{initial\_b}$ for each performance measure of interest.
   e) Repeat Steps a)-d) for all bootstrap samples, and calculate the average optimism for each performance measure. Also store the individual-level predictions $\hat{p}_{initial\_bi}$ from Step c) for each bootstrap sample ($b = 1$ to $B$). These individual-level predictions from the bootstrap model applied to the original development data can then be compared to those from the initial model, $\hat{p}_{initial\_i}$, to generate summaries and plots of the stability of individual-level predictions, to quantify the range of uncertainty around individual predictions, and the classification instability.[12]

iv) Recruit an additional $N_{new}$ patients and add them to the model development dataset; then repeat Steps (ii) and (iii) using all individuals recruited ($N_{initial} + N_{new}$). Choice of $N_{new}$ will be context specific and may be derived in a number of ways as detailed in the Discussion.

v) Repeat Step (iv), each time adding a further $N_{new}$ patients to the development dataset, until a stopping rule is met (at sample size $N_{stop}$) in terms of a desired maximum optimism in overall performance measures, or a target minimum stability in individual-level predictions and classifications (see Figure 2 for example stopping rules considered in this study)

**2.3 Stopping Rule Criteria to Determine Recommended Sample Size**

The stopping rules required in step (v) should be pre-defined and various options are considered in Figure 2. Firstly, the rules could use population-level criteria based on (optimism) in calibration and discrimination; however, in this article, our key focus is on using individual-level criteria based on precision of estimated risks and misclassification probability. Furthermore, we also consider stopping rules based on clinical utility criteria, as in many real-world applications the ultimate goal of a CPM is to support clinical decision making, where the estimated risks themselves are not of as much interest as the clinical decisions that are made using these estimates.[22-25] We consider an example risk threshold of 10% as being of clinical importance, such that estimated risks ≥10% trigger a clinician to recommend that the individual receives a form of intervention, or no treatment otherwise. See Supplementary Materials S1 and S2 for further details on the methods used here.

The final chosen sample size (accounting for all chosen stopping rule criteria simultaneously) could be influenced by random variation; there is much literature on ways to reduce this.[26, 27] In this study, for simplicity, we suggest that the criteria be met over two consecutive sample size increments before concluding that the stopping rule has been met, in line with Christodoulou et al (2021).[16] Extensions requiring consistency over three and five consecutive sample size increments were also explored, but did not change our sample size recommendations (see Supplementary Material S3 and S4).

**Figure 2.** Examples of Stopping Rule Criteria

> 1) Firstly, only relevant for unpenalised logistic regression modelling approaches with no shrinkage, we consider the population-level stability measures proposed by Christodoulou et al (2021):[16]
>
> - Bootstrap-corrected calibration slope ≥0.9
> - Mean optimism in the c-statistic ≤0.02
>
> 2) Then, for all modelling approaches now, we extend to individual-level stability of estimated risks performance measures:
>
> - Mean 95% uncertainty interval (UI) width ≤0.1, where the UI for each individual is defined by the 2.5% and 97.5% of their bootstrap predictions
> - Mean delta ≤0.05, where the delta statistic is the maximum distance between an individual's risk estimate and their lower or upper bound of the 95% uncertainty interval for their risk
>
> 3) Finally, for all modelling approaches, we also extend to clinical utility criteria.
> At the population-level, we require:
> - Expected Value of Perfect Information (EVPI) ≤0.001; where lower EVPI values indicate lower expected loss in net benefit due to uncertainty in risk predictions – see Supplementary Materials S1 and S2 for further details
>
> At the individual-level, we require:
> - Mean probability of misclassification ≤0.1; calculated as the proportion of an individual's uncertainty distribution on the opposite side of the threshold to their estimated risk (using the $B$ bootstrap risk estimates to define this uncertainty distribution, comparing the intervention decision from each of these against the 'true' risk estimate from the original model).

## 2.4 Application to the AKI Example

### *Modelling Strategies*

Given our focus was a binary outcome (development of AKI), four logistic regression model development approaches were contrasted (with or without penalisation[28] or shrinkage[29]):

- Unpenalised logistic regression, whereby all six of the core predictors mentioned in the Section 2.1 were forced into the model.

- Unpenalised logistic regression with uniform shrinkage, estimated using the heuristic shrinkage factor of Van Houwelingen and Le Cessie[30] (full details in Supplementary Material S5).
- Unpenalised logistic regression with a uniform shrinkage factor estimated using Harrell's bootstrapping approach[21] (full details in Supplementary Material S6).
- Penalised logistic regression with a Least Absolute Shrinkage and Selection Operator (LASSO) penalty term. The lasso[29] module in Stata was used, with the tuning parameter λ chosen to minimise the mean squared error on 10-fold cross-validation.

Learning curves were produced for each of these development approaches, and their required sample sizes compared across different stopping rules.

*Comparison to Riley et al Criteria*

To compare our learning curve approach to that based on the minimum recommended by the Riley et al (2019) criteria,[8] we assumed a priori that the model would provide a c-statistic of 0.78 (based on the results of Zimmerman et al (2019)[20]), and an overall AKI risk of 17.3%. Using these values, the Riley et al (2019) criteria[8] suggests we require at least 342 patients to develop our CPM (see Supplementary Material S7).

*Application of the Sequential Sample Size Calculation*

To illustrate the learning curve approach, and mirror a prospective data collection situation, we started by selecting the first $N_{initial}$=100 patients, as defined by the randomly generated recruitment ordering mentioned in Section 2.1. We use 100 as our starting point here to provide a deeper illustration of the sequential process; in practice, we recommend researchers take $N_{initial}$ to be at least the number needed to estimate the overall risk precisely (usually the least stringent component of Riley's criteria), which for this example is 220 patients (see Supplementary Material S7). We then applied the sequential method described in Section 2.2, with $B$=200 bootstrap replications. Next, we increased the sample size by selecting the subsequent $N_{new}$=100 patients from the randomly generated ordered list, then the next 100 and so on, up until a maximum of $n$=3000 individuals were reached. This allowed us to construct one learning curve for each performance measure, to show how the estimates changed as the sample size increased from 100 to 3000 in steps of 100 (see Discussion section for alternative strategies to perform re-checks of model performance). In practice, the final sample size would be that defined by the point at which the stopping rule is met ($N_{stop}$).

All analyses were performed using Stata SE version 18.0. Stata code to reproduce the example is available at https://github.com/alegha606.

All computations were performed using the University of Birmingham's BlueBEAR High Performance Computing service, provided to the University's research community. See http://www.birmingham.ac.uk/bear for more details.

## 3. Results

The results from applying the sequential sample size approach to the AKI example are now presented. We show the minimum sample size recommendations based on learning curves of population-level performance measures (Section 3.1 and 3.2), individual-level stability in predictions (Section 3.2), and clinical utility criteria (Section 3.3).

### 3.1 Population-Level Stability: Optimism in Apparent Overall Calibration and Discrimination (Unpenalised Logistic Regression Modelling Strategy)

Learning curves for population-level stability for the unpenalised logistic regression modelling strategy (without shrinkage) are displayed in Figure 3 in terms of differences in apparent and optimism-adjusted calibration and discrimination estimates. Note that the magnitude of optimism in apparent model performance can only be checked for unpenalised logistic regression, as the LASSO and uniform shrinkage approaches already adjust for optimism.

When using a small development sample size of $n$=100, large optimism in the apparent model calibration and discrimination performance is observed (bootstrap-corrected calibration slope of 0.51 and mean optimism in c-statistic of 0.10). As the sample size increases, optimism in the model calibration and discrimination performance progressively reduces. For example, at $n$=300, close to the minimum sample size recommendation by the Riley et al (2019) criteria[8] of $n$=342 (which is based on pre-defined assumptions of model performance), the bootstrap corrected calibration slope has increased to 0.78 and mean optimism in c-statistic decreased to 0.04.

### 3.2 Illustration of Safeguarding

The stopping rule criterion of bootstrap-corrected calibration slope ≥0.9 is met (and sustained over two consecutive sample size increments) at $n$=1100, while the criterion of mean optimism in c-statistic ≤0.02 is met when *n*=900. This suggests that a minimum sample size of *n*=1100 is required to achieve the chosen stability criteria for both calibration and discrimination performance for this example; this is much higher than Riley et al (2019) criteria[8] recommendation based on the assumptions made before data analysis. The key reason for the discrepancy is that the c-statistic in the full dataset that was ultimately available is much lower (0.67) than that anticipated value (0.78) assumed in the sample size calculation. Had a lower c-statistic of 0.67 been assumed in the sample size calculation, the Riley et al (2019)[8] criteria suggests a sample size of *n*=994 (see Supplementary Material S7), which is only slightly lower than the *n*=1100 observed from our sequential population-level stability criteria. This illustrates how the sequential sample size approach helps to safeguard against the consequences of inaccurate assumptions made in the sample size calculations before data collection or analysis.

### 3.3 Individual-Level Stability of Estimated Risks (All Modelling Strategies)

For stopping rules based on individual-level stability of estimated risks (mean 95% UI width ≤0.1 and mean delta ≤0.05; where the delta statistic is the maximum distance between an individual's risk estimate and their lower or upper bound of the 95% UI for their risk), large instability of estimated risks is observed for the smallest sample sizes across the various modelling approaches considered (see Figure 4 and Table 1). Then as sample size increases, this instability of estimated risks progressively decreases and eventually plateaus across higher sample sizes (as with the model calibration and discrimination).

For example, for the unpenalised logistic regression modelling strategy (i.e., without shrinkage), at a small development sample size of *n*=100, the mean 95% UI width is 0.36 and mean delta 0.23. At a higher sample size of *n*=1000, both performance statistics are closer to the desired stopping rules: the mean 95% UI width has decreased to 0.12 and the mean delta to 0.06. The mean 95% UI width ≤0.1 criteria is then met at *n*=1500, and mean delta ≤0.05 criteria at *n*=1800 ((Table 1). Thus, a minimum sample size of *n*=1800 is required to meet our individual-level stability of estimated risks criteria, which is considerably higher than the required *n*=1100 to meet our population-level criteria considering optimism in apparent calibration and discrimination.

Other modelling approaches show similar trends. For all modelling approaches, the minimum required sample size to satisfy our stopping criteria for individual-level stability of estimated risks exceeds that required for population-level stability.

**Figure 3.** Learning Curves for Population-Level Stability of Optimism in Apparent Overall Calibration and Discrimination for Unpenalised Logistic Regression Modelling Strategy

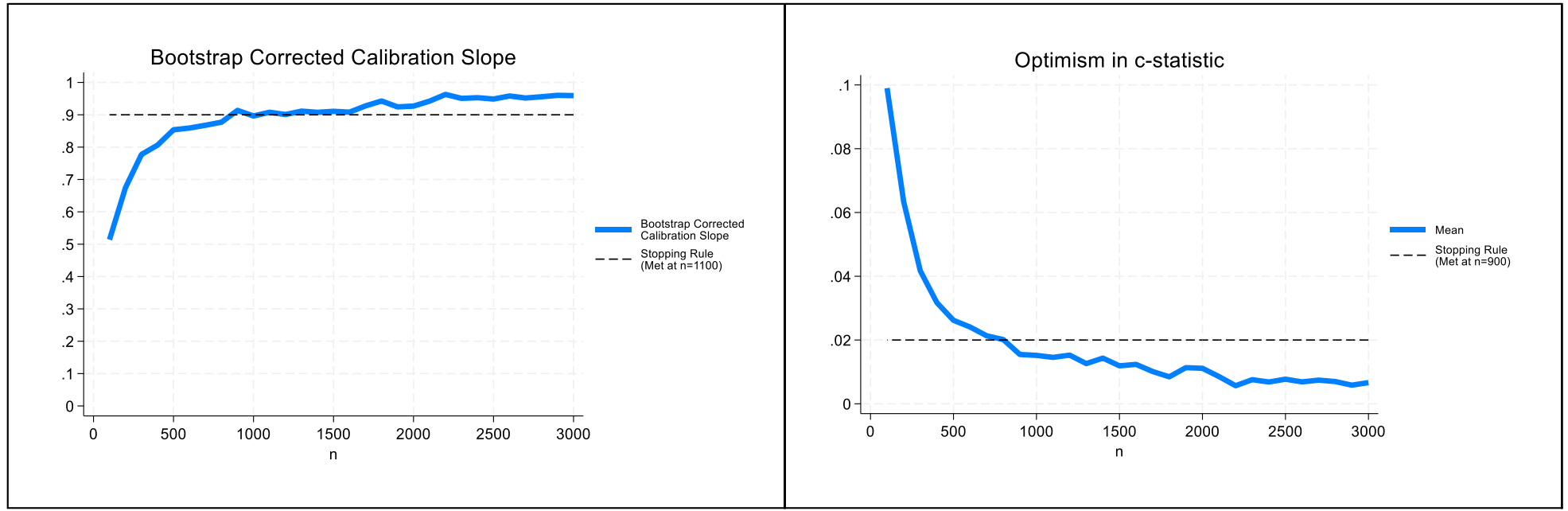

**Figure 4.** Learning Curves for Individual-Level Stability of Estimated Risks Across Different Logistic Regression Modelling Strategies (Unpenalised, Penalised, and Shrinkage Methods)

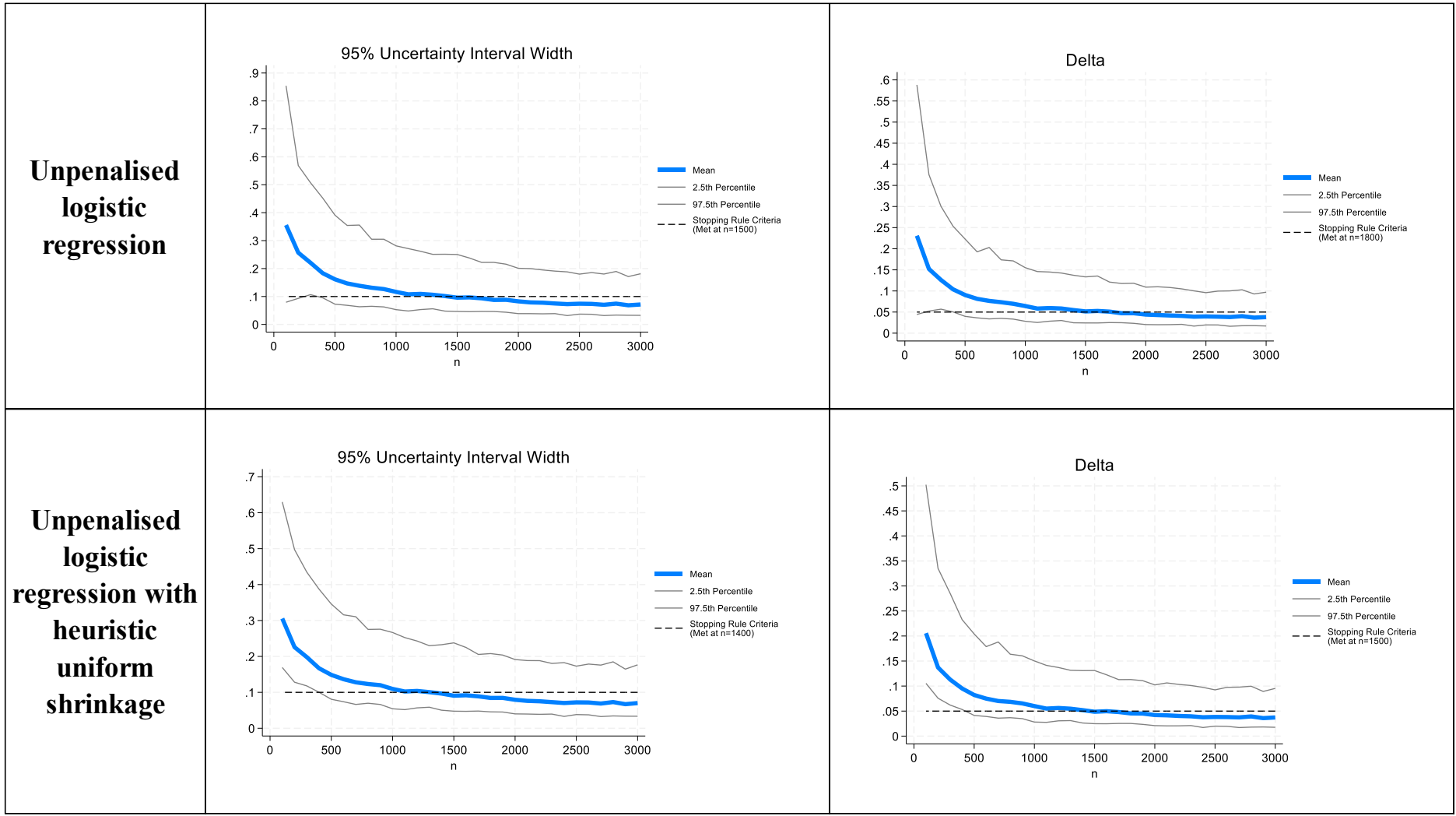

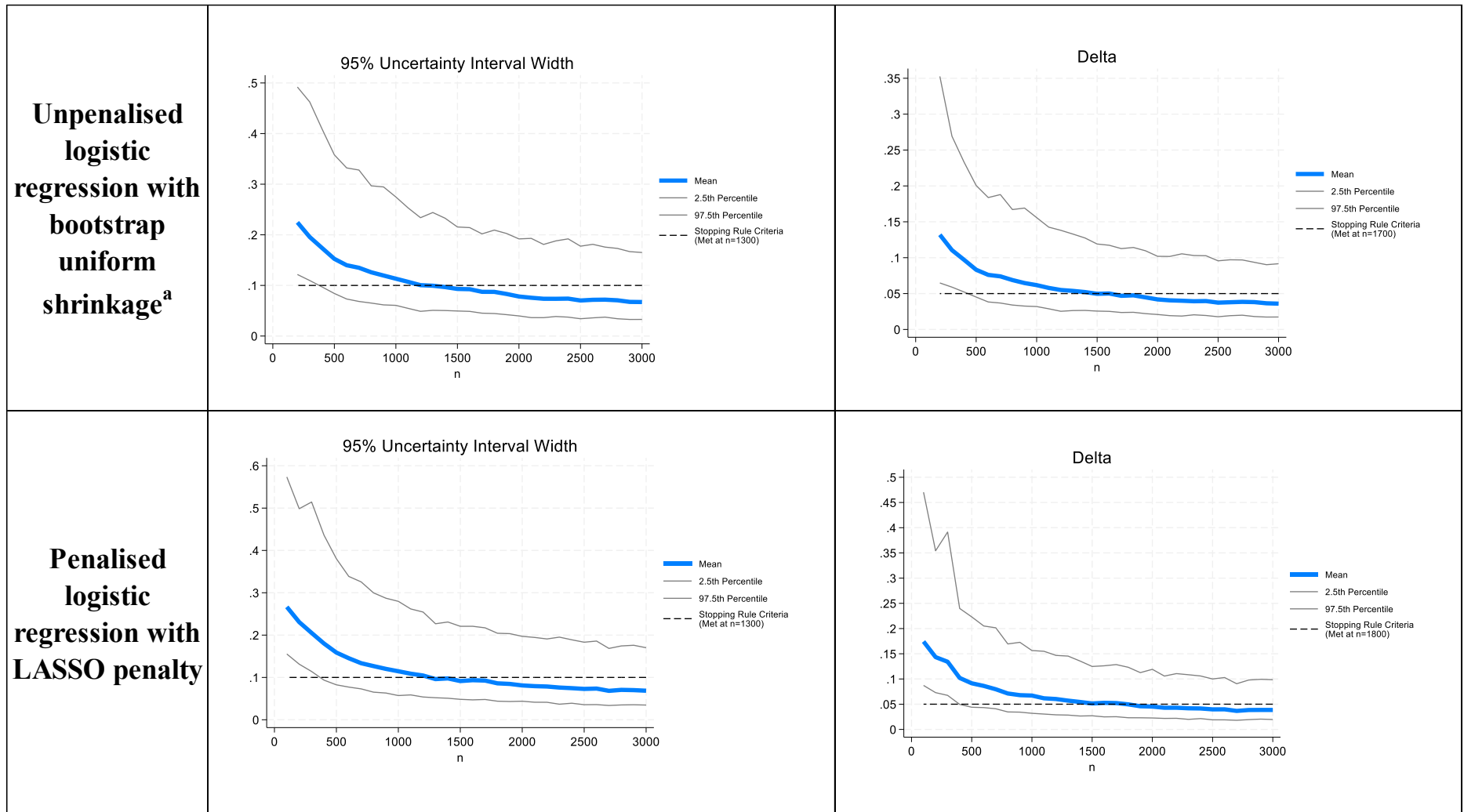

Abbreviations: LASSO = Least Absolute Shrinkage and Selection Operator.
[a] For the unpenalised logistic regression + bootstrap uniform shrinkage modelling strategy, due to model convergence issues at $n=100$, the learning curves were started from $N_{initial}=200$
Note: for individual-level performance measures, 2.5[th] and 97.5[th] percentile lines are presented, which represent the 2.5[th] and 97.5[th] values of the distribution of the given individual-level performance measure across all individuals in the current model development dataset.

**Table 1.** Summary of Minimum Required Sample Size Across Different Logistic Regression Modelling Strategies (Unpenalised, Penalised, and Shrinkage Methods) Based on Individual-Level Instability of Estimated Risks and Clinical Utility Criteria

| | | Minimum Required Sample Size | | | |
| --- | --- | --- | --- | --- | --- |
| | | Modelling Strategy | | | |
| Criteria Type | Stopping Rule | Unpenalised logistic regression | Unpenalised logistic regression with heuristic uniform shrinkage | Unpenalised logistic regression with bootstrap uniform shrinkage | Penalised logistic regression with LASSO penalty |
| Individual-Level Instability of Estimated Risks | Mean 95% UI Width $\leq 0.1$ | 1500 | 1400 | 1300 | 1300 |
| | Mean Delta $\leq 0.05$ | 1800 | 1500 | 1700 | 1800 |
| | **Overall Minimum Sample Size Recommended** | **1800** | **1500** | **1700** | **1800** |
| Clinical Utility | EVPI $\leq 0.001$ | 1600 | 1500 | 1500[a] | 1500 |
| | Mean Probability of Misclassification $\leq 0.1$ | 800 | 600 | 800 | 800 |
| | **Overall Minimum Sample Size Recommended** | **1600** | **1500** | **1500** | **1500** |

Abbreviations: LASSO = Least Absolute Shrinkage and Selection Operator; UI = Uncertainty Interval
[a] The EVPI used in the bootstrap uniform shrinkage modelling strategy is an average (the mean) of the EVPI's obtained from the 201 bootstrap shrinkage models created in each sample size iteration during the sequential process to generate a learning curve

### 3.4 Clinical Utility and Classification (All Modelling Strategies)

Instability of clinical utility and classification is high at the smallest sample sizes but reduces as sample size increases. For example, at $n$=100, for the unpenalised logistic regression model (without shrinkage), EVPI is 0.0120 (our desired criterion is ≤0.001) and mean probability of misclassification 0.22 (desired criterion ≤0.1). At a higher sample size of $n$=500, EVPI reduces to 0.0024 and mean probability of misclassification to 0.11. The EVPI learning curve (Figure 5) shows a clear trend that the improvement in net benefit from using perfect information over the current information decreases as the sample size increases; Sadatsafavi et al (2022)[24] demonstrated similar findings in their simulation study.

Clinical utility and classification instability criteria are then generally met at a slightly lower minimum sample size for each modelling strategy, compared to previous criteria based on stability of individual estimated risks (see Figure 5 and Table 1; except for the heuristic shrinkage strategy which remains at $n$=1500), and findings are largely consistent across modelling approaches. For example, the unpenalised logistic regression model (without shrinkage) requires a minimum sample size of $n$=1600 to meet classification and clinical utility criteria, rather than the previous $n$=1800 to meet individual-level stability in risk estimates.

**Figure 5.** Learning Curves for Clinical Utility Measures Across Different Logistic Regression Modelling Strategies (Unpenalised, Penalised, and Shrinkage Methods)

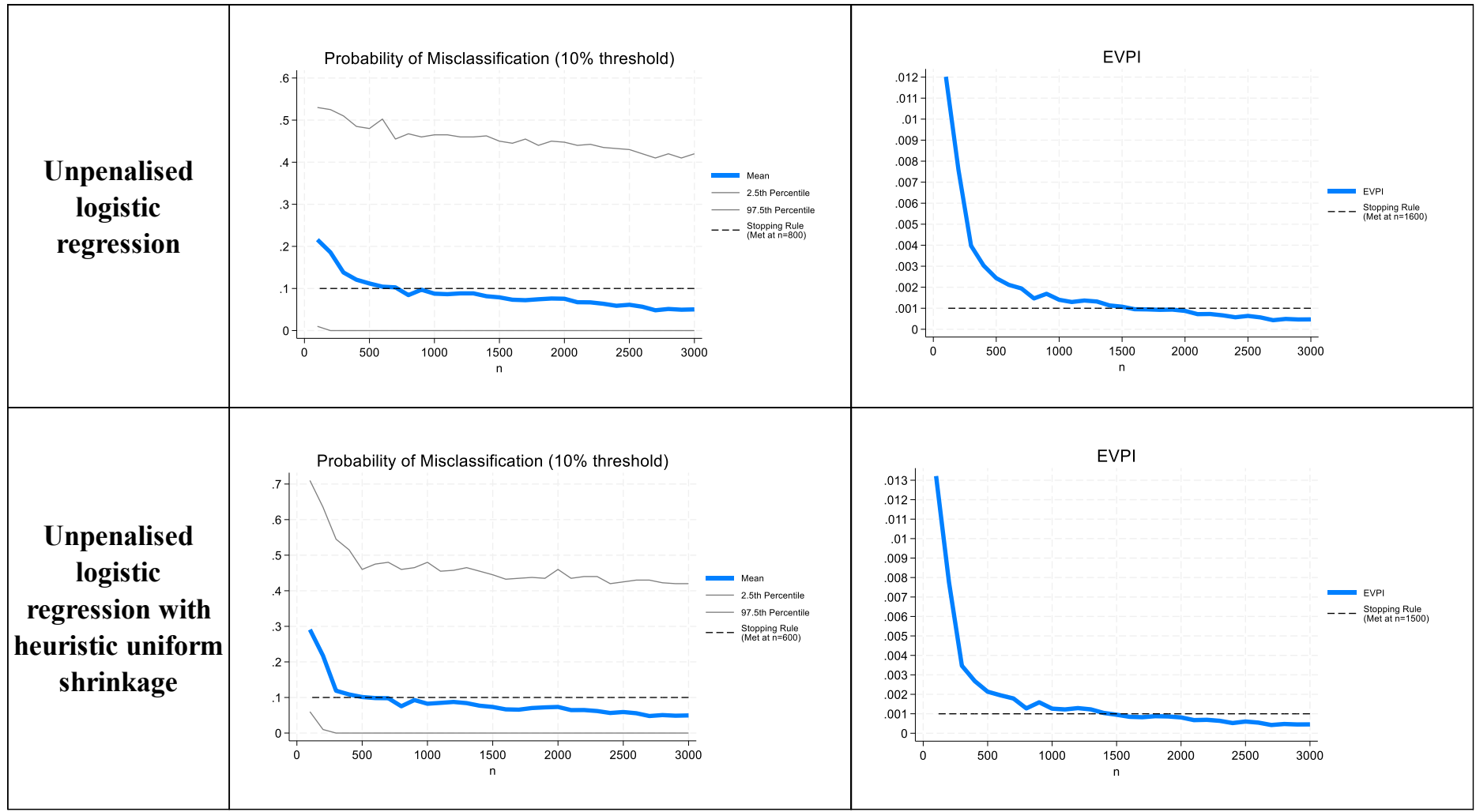

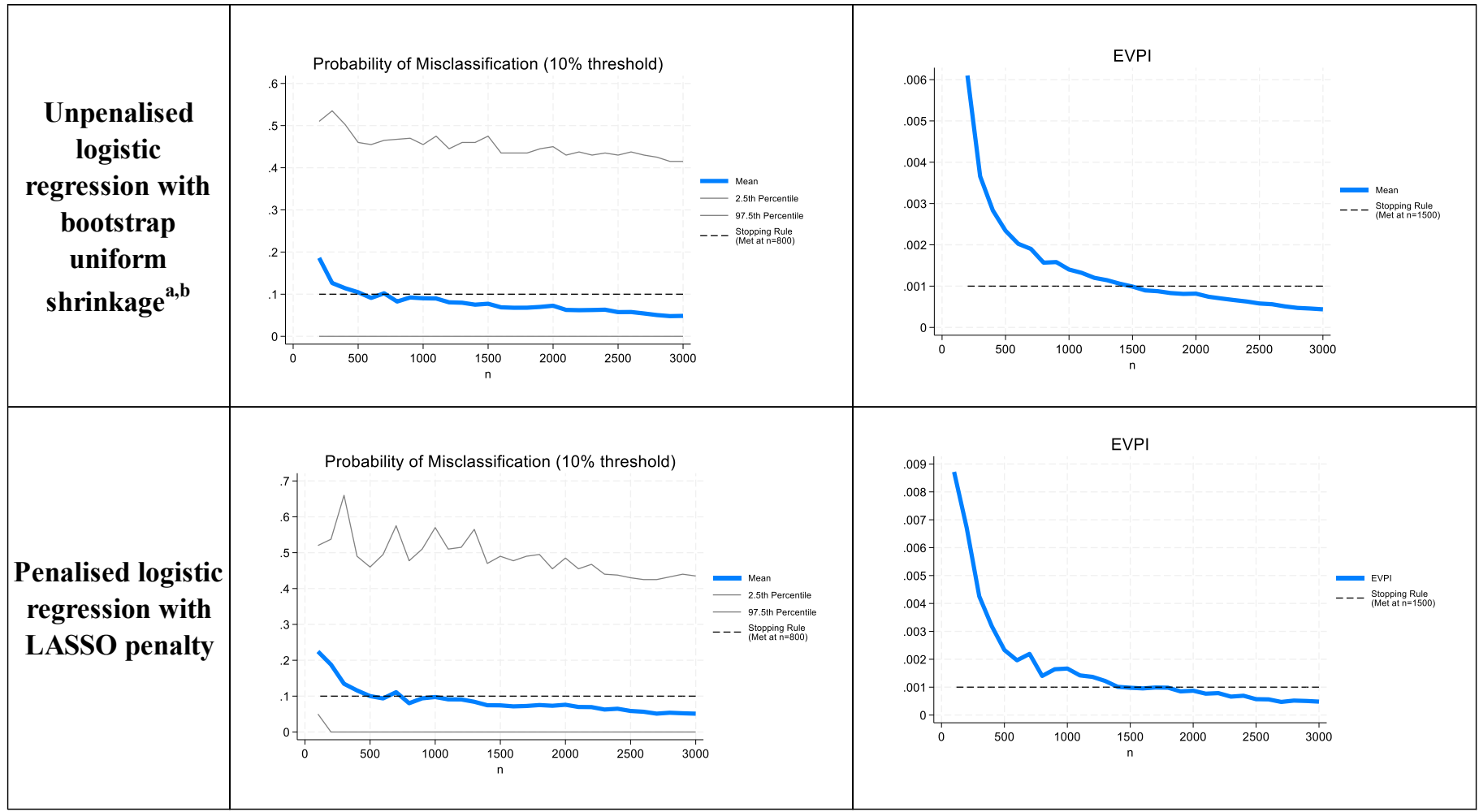

Abbreviations: EVPI = Expected Value of Perfect Information; LASSO = Least Absolute Shrinkage and Selection Operator.
[a] For the bootstrap uniform shrinkage modelling strategy, due to model convergence issues at $n$=100, the learning curves were started from $N_{initial}$=200
[b] The EVPI used in the bootstrap uniform shrinkage modelling strategy is an average (the mean) of the EVPI's obtained from the 201 bootstrap shrinkage models created in each sample size iteration during the sequential process to generate a learning curve
Note: for probability of misclassification, 2.5$^{th}$ and 97.5$^{th}$ percentile lines are presented, which represent the 2.5$^{th}$ and 97.5$^{th}$ values of the distribution of probability of misclassification across all individuals in the current model development dataset.

# 4. Discussion

We have proposed and illustrated an extension to sequential sample size calculations for CPM development studies, incorporating stopping rules based on individual-level stability in predictions and classifications alongside measures of (optimism in) calibration, discrimination and clinical utility. Also, we showed how the sequential approach can safeguard against inaccurate (e.g., overly optimistic model performance) assumptions pre-data collection. For example, assuming a c-statistic of 0.78 led to a fixed sample size recommendation of $n$=342 from the Riley et al (2019)[8] criteria. However, our sequential population-level stability criteria targeting a calibration slope ≥0.9 suggested a much higher sample size was needed ($n$=1100). It transpired that our initial c-statistic assumption was overly optimistic (in the full dataset that was ultimately available, the c-statistic was 0.67, and this value would have suggested a sample size of $n$=994 from the Riley et al (2019)[8] criteria, which is in-line with the sequential approach). Thus, in this instance, whilst the sequential method would have resulted in us recruiting more patients than initially anticipated to correct for this discrepancy in our c-statistic assumption, a more 'usable' CPM with more reliable predictions would have been developed, thereby helping to reduce research waste.

Conversely, in other situations, it may be that overly conservative model performance assumptions are made pre-data collection. For example, if we had assumed a much lower c-statistic of 0.6, then the Riley et al (2019)[8] criteria would have suggested a minimum sample size of $n$=2931. In this case, the sequential approach would have guided us to stop recruiting much earlier than anticipated (for example, at $n$=1100 if we were targeting population-level stability in terms of optimism in calibration and discrimination), which would have saved considerable time and financial resources during our study.

When extending the sequential sample size calculation approach of Christodoulou et al (2021)[16] to include stopping rules based on individual-level stability of risk estimates and clinical utility, a higher minimum sample size was required ($n$=1800 for the unpenalised logistic regression model without shrinkage, compared to $n$=1100 for stable population-level statistics). Our results align with recent research indicating that when sample sizes are predetermined before data collection, achieving stable individual-level performance often requires a significantly larger sample size than what is needed for stable population-level performance alone.[12, 14, 15, 30, 31]

Stopping rule criteria were defined by several different measures in this study, none of which are a one-stop solution to finding the 'perfect' criteria for a minimum sample size. Indeed, any relevant performance measure could be used, alongside any stopping rule criteria. It is crucial to engage with stakeholders (e.g., patients, clinicians) to determine what performance measures and associated acceptable levels of risk are relevant and acceptable to them prior to analyses, as this will be context specific. However, such conversations between model developers and stakeholders are rare.

To aid efficient implementation of our approach, rather than using a static interval of $N_{new}$ (e.g. recruitment of 50 or 100 new patients) to perform re-checks of model performance, a dynamic re-check approach may be more sensible. This could involve re-checks being based on forecasted learning curves, involving less frequent re-checks at lower sample sizes, and more frequent repeated assessments as the stopping rule criteria are approached. Such a dynamic approach allows for a more granular way of determining the final sample size, and in doing so is likely to also lead to resource savings.

Whilst a strength of our study was that we considered various types of logistic regression development approaches, we did not demonstrate our approach with a machine learning method. A random forest approach could have been tested, but we decided against this due to random forest models being known to present issues regarding model calibration;[32] further the approach already embeds bootstrapping, and so undertaking a double bootstrap for internal validation and stability checks may not be reliable, especially in smaller samples.

In summary, for model development studies carrying out prospective data collection, a sequential sample size calculation and learning curve allows researchers to dynamically monitor and identify when sufficient participants have been recruited. This safeguards against overly optimistic or conservative assumptions made for sample size calculations made in advance of data collection or analysis. Engagement with patients and other stakeholders is crucial to identify meaningful stopping rules based on key performance measures of interest. Our findings suggest that larger sample sizes are needed to achieve individual-level stability than those of traditional fixed-sample methods.

# Supplementary Material

**S1: Further Details on Methods for Determining Clinical Utility Stopping Rule Criteria**

To evaluate stability of clinical utility at the individual-level, we assess the probability of misclassification (using a stopping rule of ≤0.1); calculated as the proportion of an individual's uncertainty distribution on the opposite side of the threshold to their estimated risk (using the *B* bootstrap risk estimates to define this uncertainty distribution, comparing the intervention decision from each of these against the 'true' risk estimate from the original model).

Note that to allow us to consider the risk estimates from the model applied to the original sample as being 'true' risk estimates, Harrell's bootstrap[21] which takes a frequentist approach, is used in this article as a reasonable approximation to the Bayesian bootstrap.[24, 33] Conceptually, Harrell's bootstrap optimism correction involves viewing the sample as the population, and then using this to generate simulated samples, to which the model of interest is developed and then performance degradation in the population is assessed (i.e., in the original sample).[21] Whereas the Bayesian bootstrap views the sample as the sample, then generates random draws from the underlying population that was used to generate the sample, and the model that is then developed in the population can be interpreted as the true model, with the predicted risks from the original sample interpreted as 'true' risks.[33]

Clinical utility can also be assessed at the population-level. Sadatsafavi et al (2022 and 2023)[23, 24] describe how the Expected Value of Perfect Information (EVPI) can be used when developing a CPM, to assess the net benefit obtained by making a clinical decision on the 'correct' information (from the 'true' population model) compared to the currently available information (see Supplementary Material S2 for further description of bootstrap-based approximation for EVPI, as proposed by Sadatsafavi et al (2022)).[24] Some loss in net benefit is anticipated when using current information as opposed to perfect information; the EVPI quantifies this, with lower EVPI values indicating lower expected loss in net benefit due to uncertainty in risk predictions.

We use a stopping rule of mean EVPI ≤0.001 to inform minimum sample size recommendations. This rule was arbitrarily chosen; the case study data were used to evaluate when the expected gain in net benefit from using the proposed model from the unpenalised logistic regression (without shrinkage) was at least 99% of the expected gain from the 'true' model.

## S2: Bootstrap-Based EVPI Mechanism Used in this Study

EVPI is generated in this study using a bootstrap-based approach as proposed by Sadatsafavi et al (2022).[24] This involves adding the EVPI calculations to Harrell's bootstrap method[21] through the following addition (in bold text; Figure S1) to the learning curve generation process (outlined in Section 2.2):

**Figure S1.** Bootstrap-Based EVPI Mechanism in Learning Curve Generation Process

> a) Draw a bootstrap sample of size $N_{initial}$ (with replacement) from the initial model development dataset.
> b) Apply the modelling strategy used in Step (ii) (including the approach for any parameter tuning or data splitting) to the bootstrap sample, to obtain a bootstrap model $M_b$, where $b$ represents the bootstrap sample used to generate the model ($b$ = 1 to $B$). Calculate and store the estimated performance measures of the bootstrap model, $M_b$, when applied in bootstrap sample $b$ ($\Theta_b$).
> c) Apply the bootstrap model $M_b$ to estimate and store the risk ($\hat{p}_{initial\_bi}$) for each individual $i$ in the original development dataset ($N_{initial}$) from Step (i), and estimate and store the model performance measures $\Theta_{initial\_b}$ in that data.
> **Also calculate $NB\_All_b$, $NB\_Model_b$, and $NB\_Max_b$ (see below for equations and definitions) – these will be needed for the EVPI calculation post-bootstrap.**
> d) Calculate the optimism in model performance as the difference in the bootstrap model's apparent (in bootstrap sample) and test (in original data) performance, i.e., $\Theta_b - \Theta_{initial\_b}$ for each performance measure of interest.
> e) Repeat Steps a)-d) for all bootstrap samples, and calculate the average optimism for each performance measure. Also store the individual-level predictions $\hat{p}_{initial\_bi}$ from Step c) for each bootstrap sample ($b$ =1 to $B$). These individual-level predictions from the bootstrap model applied to the original development data can then be compared to those from the initial model, $\hat{p}_{initial\_i}$, to generate summaries and plots of the stability of individual-level predictions, to quantify the range of uncertainty around individual predictions, and the classification instability.[12]

*NB_All$_b$* represents the net benefit of administering a treatment (for example, closer monitoring of renal function) to all patients regardless of their underlying risk (for example, regardless of a patient's risk of AKI development), and is defined for bootstrap replication *b* as:

$$NB_{All_b} = mean\left[\left(\theta_{initial_b} - \frac{(1 - \theta_{initial\_b}) * z}{(1 - z)}\right)\right]$$

Where $\theta_{initial\_b}$ is the risk of AKI development obtained by fitting the bootstrap model *b* to the initial development data, and *z* represents the risk threshold, whereby if $\theta_{initial\_b} \geq z$, then a treatment is administered (*z* = 0.10 was used throughout this study).

*NB_Model$_b$* represents the net benefit of using the proposed model to decide which patients to treat, and is defined for bootstrap replication *b* as:

$$NB\_Model_b = mean\left[((\theta_{initial} \geq z) * \left(\theta_{initial\_b} - \frac{(1 - \theta_{initial\_b}) * z}{(1 - z)}\right)\right]$$

Where $\theta_{initial}$ is the risk of AKI development obtained by fitting the initial developed model to the initial development data, and the other measures are as defined above.

Finally, *NB_Max$_b$* represents the net benefit of using the 'correct' model to decide which patients to treat, and is defined for bootstrap replication *b* as:

$$NB\_Max_b = mean\left[((\theta_{initial\_b} \geq z) * \left(\theta_{initial\_b} - \frac{(1 - \theta_{initial\_b}) * z}{(1 - z)}\right)\right]$$

Where the measures are as defined above.

Then, with *NB_All$_b$*, *NB_Model$_b$*, and *NB_Max$_b$* calculated for each of *b* = 1 to *B* bootstrap replications (*B* = 200 was used throughout this study), the mean of each measure is taken across these *B* = 200 bootstrap replications:

$$ENB\_All = mean[NB\_All]$$
$$ENB\_Model = mean[NB\_Model]$$
$$ENB\_Max = mean[NB\_Max]$$

Finally, the expected value of perfect information (EVPI) is defined as:

$$EVPI = ENB\_Max - max[0, ENB\_Model, ENB\_All]$$

# S3: Minimum Sample Size Recommendations When Stopping Rule Criteria Results Met Over Three Consecutive Sample Size Increments

**Table S1.** Summary of Minimum Required Sample Size for Unpenalised Logistic Regression Modelling Strategy Based on Population-Level Instability Criteria When Stopping Rule Criteria Met Over Three Consecutive Sample Size Increments

| Population-Level Instability Criteria | Minimum Required Sample Size |
|---|---|
| Bootstrap-corrected calibration slope $\geq 0.9$ | 1100 |
| Mean optimism in c-statistic $\leq 0.02$ | 900 |
| **Overall Minimum Sample Size Recommended** | **1100** |

.

**Table S2.** Summary of Minimum Required Sample Size Across Different Modelling Strategies Based on Individual-Level Instability Criteria When Stopping Rule Criteria Met Over Three Consecutive Sample Size Increments

| | | Minimum Required Sample Size | | | |
|---|---|---|---|---|---|
| | | Modelling Strategy | | | |
| Criteria Type | Stopping Rule | Unpenalised logistic regression | Unpenalised logistic regression with heuristic uniform shrinkage | Unpenalised logistic regression with bootstrap uniform shrinkage | Penalised logistic regression with LASSO penalty |
| Individual-Level Instability of Estimated Risks | Mean 95% UI Width ≤ 0.1 | 1500 | 1400 | 1300 | 1300 |
| | Mean Delta ≤ 0.05 | 1800 | 1500 | 1700 | 1800 |
| | **Overall Minimum Sample Size Recommended** | **1800** | **1500** | **1700** | **1800** |
| Clinical Utility | EVPI ≤ 0.001 | 1600 | 1500 | 1500[a] | 1500 |
| | Mean Probability of Misclassification ≤ 0.1 | 800 | 600 | 800 | 800 |
| | **Overall Minimum Sample Size Recommended** | **1600** | **1500** | **1500** | **1500** |

Abbreviations: LASSO = Least Absolute Shrinkage and Selection Operator; UI = Uncertainty Interval

[a] The EVPI used in the bootstrap uniform shrinkage modelling strategy is an average (the mean) of the EVPI's obtained from the 201 bootstrap shrinkage models created in each sample size iteration during the sequential process to generate a learning curve

**S4: Minimum Sample Size Recommendations When Stopping Rule Criteria Results Met Over Five Consecutive Sample Size Increments**

**Table S3.** Summary of Minimum Required Sample Size for Unpenalised Logistic Regression Modelling Strategy Based on Population-Level Instability Criteria When Stopping Rule Criteria Met Over Five Consecutive Sample Size Increments

| Population-Level Instability Criteria | Minimum Required Sample Size |
|---|---|
| Bootstrap-corrected calibration slope ≥0.9 | 1100 |
| Mean optimism in c-statistic ≤0.02 | 900 |
| **Overall Minimum Sample Size Recommended** | **1100** |

**Table S4.** Summary of Minimum Required Sample Size Across Different Modelling Strategies Based on Individual-Level Instability Criteria When Stopping Rule Criteria Met Over Five Consecutive Sample Size Increments

| | | Minimum Required Sample Size | | | |
|---|---|---|---|---|---|
| | | **Modelling Strategy** | | | |
| **Criteria Type** | **Stopping Rule** | **Unpenalised logistic regression** | **Unpenalised logistic regression with heuristic uniform shrinkage** | **Unpenalised logistic regression with bootstrap uniform shrinkage** | **Penalised logistic regression with LASSO penalty** |
| **Individual-Level Instability of Estimated Risks** | Mean 95% UI Width ≤ 0.1 | 1500 | 1400 | 1300 | 1300 |
| | Mean Delta ≤ 0.05 | 1800 | 1500 | 1700 | 1800 |
| | **Overall Minimum Sample Size Recommended** | **1800** | **1500** | **1700** | **1800** |
| **Clinical Utility** | EVPI ≤ 0.001 | 1600 | 1500 | 1500[a] | 1500 |
| | Mean Probability of Misclassification ≤ 0.1 | 800 | 600 | 800 | 800 |
| | **Overall Minimum Sample Size Recommended** | **1600** | **1500** | **1500** | **1500** |

Abbreviations: LASSO = Least Absolute Shrinkage and Selection Operator; UI = Uncertainty Interval
[a] The EVPI used in the bootstrap uniform shrinkage modelling strategy is an average (the mean) of the EVPI's obtained from the 201 bootstrap shrinkage models created in each sample size iteration during the sequential process to generate a learning curve

## S5: Further Details on the Heuristic Shrinkage Factor

In the heuristic shrinkage modelling strategy, a uniform shrinkage factor is applied to the beta coefficients of our logistic regression base-case model (on the natural logarithmic scale) and the model intercept is re-estimated.

Consider that the base-case logistic regression model in this study can be expressed as:

$$ln\left(\frac{p}{1-p}\right) = \alpha + \beta_1 * bicarbonate + \beta_2 * creatinine + \beta_3 * haemoglobin + \beta_4 * blood\ urea\ nitrogen + \beta_5 * potassium + \beta_6 * systolic\ blood\ pressure$$

Where:

- $p$ is the probability of development of AKI,

- $\alpha$ is the intercept, the log-odds of AKI development when all of the six continuous predictors are 0,

- and $\beta_1$- $\beta_6$ the log-odds ratio of AKI development for a one unit increase in bicarbonate, creatinine, haemoglobin, blood urea nitrogen, potassium, and systolic blood pressure, respectively (whilst each of the other predictors are held constant)

Then, in the heuristic shrinkage model, a shrinkage factor, *H*, is applied to produce optimism-adjusted beta coefficients such that:

$$ln\left(\frac{p}{1-p}\right) = \alpha^* + (H * [\beta_1 * bicarbonate + \beta_2 * creatinine + \beta_3 * haemoglobin + \beta_4 * blood\ urea\ nitrogen + \beta_5 * potassium + \beta_6 * systolic\ blood\ pressure])$$

Where:

- $\alpha^*$ is the re-estimated intercept (after having applied the heuristic shrinkage factor *H*),

- and the heuristic shrinkage factor, *H*, is as defined by Van Houwelingen and Le Cessie[29] as:

$$H = \frac{model\ \chi^2 - df}{model\ \chi^2}$$

With *model* $\chi^2$ representing the chi-squared statistic (likelihood ratio) for the model, and *df* the total degrees of freedom for the predictors.

**S6: Bootstrap Uniform Shrinkage Modelling Strategy**

In the bootstrap uniform shrinkage modelling strategy, for a given iteration of the learning curve generation process (see Section 2.2), an initial development sample of fixed size is first chosen without replacement from the $n$=20,413 patients available in the MIMIC dataset. Then Harrell's bootstrap approach[21] is applied to produce a bootstrap-corrected calibration slope that is used as a shrinkage factor to produce a bootstrap penalised model (the uniform shrinkage process is as detailed in Supplementary Material S5, but with the bootstrap-corrected calibration slope used for shrinkage instead of the heuristic shrinkage factor), this shrunken model then gives a corresponding 'original' risk estimate.

To generate instability risk estimates for a given individual (see Section 2.3 for individual-level performance measures considered in this study), $B$=200 bootstrap samples are then drawn from the original development data (to produce $B$=200 'new' development datasets), then Harrell's bootstrap[21] is applied to produce a penalised model for each of these $B$=200 'new' development datasets, and finally these penalised models are applied back to the original development data to give 200 instability risk estimates for each individual.

## S7: Stata Code for Riley et al Criteria

The Stata package *pmsampsize*[10] was used to perform a pre-data collection fixed approach to minimum sample size calculation from the method of Riley et al (2019),[8] using the following information:

- Prevalence of outcome (development of AKI in this study), 17.3%
- C-statistic of 0.78 from the Zimmerman et al (2019)[20] study
- Six predictor parameters used, based on those from Zimmerman et al (2019)[20]

The corresponding Stata code is:

*pmsampsize, type(b) cstatistic(0.78) parameters(6) prevalence(0.173)*

This suggests that at least 342 patients (with 60 events) are needed for developing this CPM with six predictor parameters.

This code output also shows us that if we are only interested in the criterion that requires precise estimation of the average outcome risk in the population (rather than the full Riley et al (2019)[8] criteria), then at least 220 patients are needed.

Note that when amending the c-statistic used in this minimum sample size calculation to that obtained by applying the model of interest to the full MIMIC-III dataset available of $n$=20,413 patients (rather than using the c-statistic from the clinically relevant Zimmerman et al (2019)[20] study), the corresponding Stata code is now:

*pmsampsize, type(b) cstatistic(**0.67**) parameters(6) prevalence(0.173)*

Hence using the c-statistic from the available data rather than making an assumption to approximate the c-statistic suggests that at least 994 patients (with 172 events) are needed for developing this CPM with six predictor parameters.